\documentclass[aps,pra,twocolumn,longbibliography,superscriptaddress,final,floatfix, 11pt,tightenlines]{revtex4-2}

\usepackage{amsfonts}
\usepackage{amsmath}
\usepackage{amssymb}
\usepackage{mhchem}
\usepackage{subfigure}
\usepackage{latexsym}
\usepackage[export]{adjustbox}
\usepackage{array}
\newcolumntype{P}[1]{>{\centering\arraybackslash}p{#1}}
\newcolumntype{M}[1]{>{\centering\arraybackslash}m{#1}}
\usepackage{caption}
\usepackage[colorlinks,citecolor=red,urlcolor=blue,bookmarks=false,hypertexnames=true]{hyperref} 
\usepackage[usenames, dvipsnames]{color}
\usepackage[svgnames]{xcolor}
\usepackage{color}
\usepackage{bm}
\usepackage{relsize}
\usepackage{mathtools}
\usepackage{float}
\usepackage{subfloat}
\usepackage{dcolumn}
\usepackage{epsfig}
\usepackage{graphicx}
\usepackage{multirow}
\usepackage{makecell}
\usepackage{mathtools}
\usepackage[bbgreekl]{mathbbol}

\usepackage{footnote}
\usepackage{balance}

\usepackage{mathrsfs}
\begin{document}

\title{Decoherence of a  dissipative Brownian charged magneto-anharmonic oscillator: an information theoretic approach}
\author{Suraka Bhattacharjee}
\thanks{Corresponding Author \\
Email: surakabhatta@gmail.com}
\affiliation{SASTRA Deemed University, Thirumalaisamudram, Thanjavur-613401, India} 
\author{Koushik Mandal}
\affiliation{School of Applied Science and Humanities, Haldia Institute of Technology, Haldia-721657, India}    
\author{Supurna Sinha}
\affiliation{Raman Research Institute, Bangalore-560080, India}
\date{\today}
\begin{abstract}
We study the decoherence of an anisotropic anharmonic oscillator in a magnetic field, coupled to a bath of harmonic oscillators at high and low temperatures. We solve the anharmonic oscillator problem using perturbative techniques and derive the non-Markovian master equation in the weak coupling limit. The anharmonicity parameter $\alpha$ enhances decoherence due to the deconfining effect of anharmonicity. The oscillatory nature of the time evolution of heating function indicates information backflow. The von-Neumann entropy is also calculated for the system, which increases with $\alpha$, consistent with the deconfining effect noted in the decoherence analysis. We have also proposed a cold ion experimental set up for 
testing our theoretical predictions. The study is of relevance to the domain of quantum technology where decoherence significantly affects the performance of a quantum computer.
\end{abstract}
\maketitle
\section{Introduction}
Typically a real quantum system is not isolated. It is 
coupled to an environment. The system-bath interaction which stems from this coupling, entangles the quantum system to a large number of environmental degrees of freedom. This coupling leads to an environment-induced decoherence which in turn results in a loss of information from the system to the surroundings, giving rise to a quantum-to-classical transition \cite{Zurek1991,Schlosshauer2007}. \\
In \cite{Zeh1970}, the author identified quantum decoherence as a signature of quantum-to-classical transition in real quantum systems coupled to an external environment. Subsequently, there has been considerable progress and research in the field of decoherence, aiming at reducing the loss of information from the system to the surroundings \cite{Zeh1970,Zurek1981,Zurek1982,Zurek2003,Schlosshauer2004,Bacciagaluppi2003,Schlosshauer2007,Science2008,PRB2023,JPCM2025}. Recently, the control of decoherence with plasmonics has also played a vital role in realizing photons as carriers in quantum information processing via photon-photon interactions \cite{Bogdanov2019}. In \cite{Chen2024}, the authors have studied the separability transitions in topological states induced by local decoherence. Decoherence was also observed in several experiments including matter-wave interferometry, cavity QED, superconducting systems, ion traps, quantum mechanical resonators and in Bose-Einstein condensates (BEC) \cite{Schneider1998,Schneider1999,Hanson2007,Aspelmeyer2014,Burkard2004,Ithier2005,Burnett2019,Kuang1999}. In addition, computer simulations of solid state spin qubits paved the way for controlling decoherence and designing next general quantum technologies \cite{Onizhuk2025}. Furthermore, over the years, theoretical modeling of decoherence in quantum open systems has evolved from toy models to 
more realistic ones closer to complex systems observed in nature \cite{Schlosshauer2007}. In some of our earlier works, we have analyzed the loss of coherence in harmonically oscillating quantum Brownian particles in the presence of a magnetic field and coupled to various types of heat baths \cite{decoherence1,decoherence2}. The coupling with the environment was also generalized to both position and momentum coordinate couplings, which mimic realistic quantum systems \cite{decoherence2,suraka2}.  \\
 A relatively less explored domain is the role of anharmonicity in the decoherence of a
quantum system coupled to an environment. Typically one considers the subsystem to be a harmonic oscillator. However, in a realistic system there can be deviations from a strictly harmonic oscillator potential and anharmonicity can lead to decofinement of the trapped particle. One therefore expects 
an anharmonic system to be more susceptible to the decohering effect of the environment. This is the motivation behind our present study. Thus our goal is to understand the role of anharmonicity in the 
destruction of coherence in a quantum system coupled to an environment.\\
The present work is focused on the study of decoherence of a charged anharmonic oscillator in the presence of a magnetic field coupled
to a dissipative environment characterized by an Ohmic spectral density function. The Markovian dynamics of simple anharmonic Brownian motion models were studied earlier both theoretically and experimentally \cite{chung2013lectures}. The semi-classical non-Markovian dynamics of a Brownian particle was numerically analyzed in a Morse potential and compared with the results obtained from the standard Caldeira–Leggett master equation \cite{Koch2010}. The decoherence of anharmonic oscillators in the presence of Morse potentials has been further addressed in some theoretical works in the context of photon or phonon modes in thermal equilibrium \cite{Elran2004,Foldi2003}. However, the analytical study of decoherence using a master equation involves solving the system equations of motion in terms of the initial coordinates \cite{suraka1,decoherence1,Schlosshauer2019quantum}. The method used for exactly solving an anhamonic oscillator for quartic anharmonicity has been extended to higher order even powers in the 
anharmonicity parameter as well \cite{Flessas1981}. In addition to these exact methods, various perturbative techniques are applied to solve for different types of anharmonicities, when the strength of the anharmonic potential is very small compared to its harmonic counterpart \cite{Robinnet19977}. Here, we have used a perturbative method to solve the dynamics of an anharmonic oscillator in the presence of a weak magnetic field. We have taken into consideration non-Markovian effects, which account for environmental memory, that significantly influences the evolution of the system. In most of the earlier works, open quantum systems (OQS) have been mainly studied using the Caldeira-Leggett approach within the Markovian approximation\cite{chung2013lectures,diosi1993}. However, in the presence of non-Markovianity an OQS where an anharmonic potential 
is coupled to an environment has the potential of emerging as an active area of research both theoretically as well as experimentally\cite{exptanharmonic,dupont2018experimental}.  Thus non-Markovianity adds an important dimension to our study enabling us to explore non-trivial memory effects appearing in the quantum to classical transition via decoherence. In this paper, we have proposed an experimental set up involving a Penning ion trap for trapping charged particles and counter-propagating laser beams to create an optical molasses in the presence of an external magnetic field. We make suitable approximations 
to render our theoretical calculations and predictions tractable. 
Our aim is to highlight the combined effects of these relatively less explored features on the decoherence of a quantum system. 
 We also study the time evolution of the heating function $F_h(t)$, which involves the off-diagonal elements of the reduced density matrix and thus is central to decoherence studies  \cite{malay2,Malay2015,Paavola}.
Furthermore, we calculate the von Neumann entropy  ($S_{VN}$) from the reduced density matrix using the Wigner function and related Weyl transformation and thus obtain a measure of the system-environment entanglement \cite{weiderpass2020,bellac2004}.  
The von Neumann entropy is zero for pure states. Its non-zero value signifies a mixed state and the extent of statistical uncertainty even at zero temperature. 
The nonzero value of $S_{VN}$ at zero temperature is a consequence of entanglement 
of the subsystem and the environment. The study of $S_{VN}$ is crucial in the domain of quantum information.
Thus we go beyond a master equation analysis of studying quantum dephasing via the 
time evolution of the off diagonal elements of the reduced density matrix and study the von Neumann entropy $S_{VN}$, to view the problem through a quantum information theoretic lens. 
\\
The paper is outlined as follows: In section $2$, we develop the quantum Langevin equation (QLE) for a charged quantum particle in an anisotropic anharmonic potential, in the presence of a magnetic field and coupled to an environment of harmonic oscillators. A perturbative approach has been introduced in a subsection to get the solution for the particle in the $2D$ plane. The non-Markovian master equation is set up for the particle and using this, the dynamics of the reduced density matrix (RDM) is derived in section $3$. We calculate the von Neumann entropy for the particle in section $4$. In section $5$, we have proposed a suitable experimental set up to test our theoretical predictions. The decay of coherence of the RDM and the evolution of the function $h(t)$ 
 and its time integral, the heating function $F_H(t)$ with time are presented and the results are discussed in section $6$.    
\section{Quantum Langevin Equation}
We consider a charged quantum Brownian particle trapped in an anharmonic potential in the presence of an external magnetic field. The magnetic field is along the $z$-axis and the motion of the charged particle is confined to the two-dimensional $x-y$ plane.  Here we solve the equations of motion for the problem and determine the system operators which are essential for deriving the dynamics of the reduced density matrix and the decoherence process. \\
The Hamiltonian
for this system coupled to a bath of harmonic oscillators
is given by:
\begin{equation}
    H = H_{S} + H_{E} + H_{SE}
\end{equation}
with
\begin{align}
    H_{S} &= \frac{1}{2m}\left(p-\frac{eA}{c}\right)^{2} + V(x,y) \label{syshamiltonian}\\
     H_{E}&=\sum_{j} \frac{p_{j}^{2}}{2m_{j}} +\frac{1}{2}m_{j}\omega_{j}^{2}q_{j}^{2} \label{envhamiltonian}
\end{align}
where $H_{S}$ and $H_{E}$ are respectively the Hamiltonian of the system and the environment.  $A$ $[(By/2,-Bx/2,0)]$ is the vector potential pertaining to the 
applied magnetic field $B$ and $p$, $x$, $y$, $m$ are respectively the momentum, position coordinates and mass of the particle.
$p_{j}$, $q_{j}$, $m_{j}$ and $\omega_{j}$ are the
momentum, the position coordinates, mass and frequency
of the $j$-th bath oscillator. \\
We consider the following potential:
\begin{equation}
    V(x,y)= m\omega_0^2\left[\frac{1}{2}(x^{2}+y^{2}) -\alpha x^{3} \right]
\end{equation}
where $\omega_0$ is the frequency of the harmonic oscillator and $\alpha$ is the anharmonicity parameter in the $x$ direction. Thus we have introduced an anisotropic anharmonicity in the potential as has been reported in some theoretical and experimental studies \cite{Joanna2013}.\\
We consider position-position coupling and the corresponding particle-bath interaction can be modeled in the following form: 
\begin{equation}
    H_{SE} = x \otimes \sum_{j} c_{j}q_{jx} + y \otimes \sum_{j} c_{j}q_{jy} \label{sys-envhamiltonian}
\end{equation}
where the coupled $x$ and $y$ coordinates are monitored by the environment. The system is linearly coupled to the environment via the position coordinate and $c_{j}$ is the coupling constant.  $q_{jx}$ and $q_{jy}$ are respectively the $x$ and $y$ components of 
$q_{j}$.\\
We solve the system equations of motion using a perturbative method to compute the extent of decoherence, determined by the decay of the off-diagonal elements of the reduced density matrix. 
\subsection{Perturbative Solution}
Here we consider the anharmonicity in the $x$-direction. We use perturbative tools and solve the coupled differential equations derived from the system Hamiltonian (Eq.(\ref{syshamiltonian})) \cite{Robinnet19977}.\\
\begin{align}
 &\Ddot{x}(t) + \omega_0^2 x(t) + 3 \alpha \omega_0^2 x^2(t)-\omega_c \dot{y}(t)=0 \label{xeq1}\\
 &\Ddot{y}(t) + \omega_0^2 y(t) + \omega_c \dot{x}(t)=0 \label{yeq1}
\end{align}
where, $\omega_c$ is the cyclotron frequency ($\omega_c=\frac{eB}{m}$).\\
Let us consider the perturbative solution in $x$ and $y$ in the following form \cite{Robinnet19977}:
\begin{align}
    x(t)= x_{0}(t)+X \alpha x_{1}(t)+(X \alpha)^{2}x_{2}(t)+...
    \label{p_sol_1}
    \\
    y(t)= y_{0}(t)+Y\alpha y_{1}(t)+ (Y\alpha)^{2}y_{2}(t)+...
    \label{p_sol_2}
\end{align}
$x_0(t)$ and $y_0(t)$ are the solutions of the coupled Langevin equations for a harmonic oscillator in the presence of a magnetic field, as presented in our earlier works \cite{decoherence1,decoherence2}:\\
As $x_0$ and $y_0$ are the exact solutions to the harmonic oscillator problem, using Eqs.(\ref{p_sol_1}-\ref{p_sol_2}) we get from Eqs.(\ref{xeq1}-\ref{yeq1}), neglecting terms higher than first order in the anharmonicity parameter $\alpha$:
\begin{align}
 X\Ddot{x_1}(t) + X\omega_0^{2} x_1(t) -3\omega_0^{2} x_0^2(t)-Y\omega_{c}\dot{y_1}(t) =0 \label{perturb1}\\
    Y\Ddot{y_1}(t)+ Y\omega_0^{2} y_1(t) +X\omega_{c}\dot{x_1}(t)= 0  \label{perturb2}
\end{align}
where, $X$ and $Y$ are the initial $x$ and $y$ coordinates of the particle respectively.\\
We solve the coupled equations (Eqs.[\ref{perturb1}-\ref{perturb2}]) in the small $\omega_c$  ($\omega_c << \omega_0$) limit, using standard techniques and get $x_1(t)$ and $y_1(t)$ as:
\begin{align}
   & x_{1}(t)= f_{0} (t) X + f_{1} (t) Y +  \notag \\
    &f_{2} (t) (Y^{2}/X) +\frac{\sin(\omega_0 t)}{\omega_0} V_x \label{x1sol}\\
     &y_{1}(t)= Y \cos(\omega_0  t) +\frac{V_y}{\omega_0} \sin(\omega_0 t) \label{y1sol}
\end{align}
In Eq.(\ref{x1sol}), $f_{0} (t), f_{1} (t)$ and $f_{2} (t)$ represent time-varying functions that also depend on the constants $\omega_{0}$ and $\omega_{c}$ (see Appendix-A).\\
Using the expression for $x_1$ and $y_1$, one can derive the perturbative solution as:
\begin{align}
    &x(t)= x_0(t)+ \alpha (f_0(t) X^2 + f_1(t) XY + \notag \\
    &f_2(t) Y^2+\frac{\sin(\omega_0 t)}{\omega_0} X V_x) \label{solutionx}\\
   & y(t) =y_0(t) + \alpha \cos (\omega_0 t) Y^2 + \frac{\sin(\omega_0 t)}{\omega_0} Y V_y \label{solutiony}
\end{align}
where,  $x_0(t)$ and $y_0(t)$ in the small $\omega_c$ limit are given by \cite{decoherence1,decoherence2}:
\begin{widetext}
\begin{align}
    x_0(t)=&\frac{1}{4 \omega_0 \omega_c}\bigg[\bigg \lbrace 2 \omega_0 \omega_c \cos(At)+2 \omega_0\omega_c \cos(Bt) \bigg \rbrace X +\bigg \lbrace 2i\sqrt{2} \omega_0^2 \omega_c \left(\frac{\sin(A t)}{A}-\frac{\sin(B t)}{B} \right) \bigg \rbrace Y+ \notag \\
    &\bigg \lbrace \frac{2i\sqrt{2}\omega_0 \omega_c}{A}\sin(A t)+ \frac{2\sqrt{2}\omega_0 \omega_c}{B} \sin(B t) \bigg \rbrace V_x +\bigg \lbrace 2\omega_c \left( -\cos(A t)+\cos(B t)\right) \bigg \rbrace V_y\bigg] \label{couplx}\\
   y_0(t)=&\frac{1}{4 \omega_0 \omega_c }\bigg[\bigg \lbrace 2\omega_0 \omega_c \cos(A t)+ 2\omega_0 \omega_c \cos(B t) \bigg \rbrace Y -\bigg \lbrace 2i\sqrt{2} \omega_0^2 \omega_c \left(\frac{\sin(A t)}{A}-\frac{\sin(B t)}{B} \right) \bigg \rbrace X+ \notag \\
   &\bigg \lbrace \frac{2i\sqrt{2} \omega_0 \omega_c}{A}\sin(A t)+ \frac{2i\sqrt{2} \omega_0 \omega_c}{B} \sin(B t) \bigg \rbrace V_y -\bigg \lbrace 2\omega_c \left( -\cos(A t)+\cos(B t)\right) \bigg \rbrace V_x\bigg] \label{couply}
\end{align}
\end{widetext}
   Here, $V_x$ and $V_y$ represent the initial velocity of the Brownian particle in the $x$ and $y$ direction respectively. The coefficients $A$ and $B$ are given by: $ A=\sqrt{\omega_0(\omega_0+\omega_c)}$  and $B=\sqrt{\omega_0(\omega_0-\omega_c)}$.
\section{Non-Markovian Master Equation: Dynamics of the reduced density matrix}
The Liouville-von Neumann equation for the total density operator in the interaction picture is given by \cite{Schlosshauer2007,Breuer2007}:
\begin{align}
    \frac{\partial}{\partial t} \rho^{(I)}(t)= \frac{1}{\hbar}\left[H_{int}(t),\rho^{(I)}(t) \right] \label{Neumann}
\end{align}
The non-Markovian master equation can be derived from the Liouville-von Neumann equation \cite{Redfield,Blum}:
\begin{align}
    \frac{\partial}{\partial t} \rho_s(t)&=-\frac{i}{\hbar}\left[H_s,\rho_s(t) \right]- \notag \\
    &\frac{1}{\hbar^2}\left \lbrace \left[S_{\alpha},B_{\alpha}\rho_s(t) \right]  + \left[\rho_s(t)C_{\alpha},S_{\alpha} \right] \right \rbrace \label{Born-Markovgen}
\end{align}

where,
\begin{align}
    B_\alpha=\int_0^t d\tau \sum_\beta C_{\alpha \beta}(\tau) S_\beta ^{(I)}(-\tau) \label{Balphagen}\\
    C_\alpha=\int_0^t d\tau \sum_\beta C_{\beta \alpha}(-\tau) S_\beta ^{(I)}(-\tau) \label{Calphagen}
\end{align}

Operators with superscript $(I)$ pertain to the interaction picture and the ones without a superscript represent Schr\"odinger picture operators. 
Here $S_\alpha ^{(I)}(-\tau)$ denotes the system operator $S_\alpha$ in the 
interaction picture.  $C_{\alpha \beta}(\tau) = \langle\hat{E}_{\alpha}(\tau)\hat{E}_{\beta}\rangle_{\rho_E}$ is the environment self-correlation function pertaining to the operator $\hat{E}$ measured on the environment as a result of the system-environment interaction where the average is taken over the initial state $\hat{\rho}_E$ of the environment. It quantifies how much the result of the measurement of an observable is correlated with the result of 
a measurement of the same observable at a time $\tau (\tau=t-t')$ later. Thus, this function quantifies the extent to which the environment retains information over time about its interaction with the system. So the Markov approximation corresponds to the assumption of a rapid decay of these environment self correlation functions relative to the timescale set by the evolution of the system.
Further, the self correlation function $C_{\alpha \beta}(\tau)$ is peaked at $\tau=0$ for the Markovian case, whereas, the peak broadens as the non-Markovian limit is achieved, signifying the presence of the finite memory effect \cite{Schlosshauer2007}. However, the Born-Approximation is still considered to be valid, as the interaction between the system and the bath is weak such that the system-environment density matrix is approximated as the tensor product of the system density matrix and the environmental density matrix ($\rho_{SE}(t)\approx \rho_S(t) \otimes \rho_E$) \cite{Schlosshauer2007,Schlosshauer2019quantum}:\\
Using the model for an anharmonic oscillator in a magnetic field Eq.(\ref{Born-Markovgen}) reduces to \cite{Schlosshauer2007}:
\begin{align}
    &\frac{\partial \rho_s (t)}{\partial t}=-\frac{i}{\hbar}\left[H_s, \rho_s(t) \right]- \notag \\
    &\frac{1}{\hbar}\int_0^t d\tau \bigg \lbrace \nu(\tau)\left[ x,[x(-\tau),\rho_s(t)\right]]-\notag \\
    & i \eta(\tau))\left[ x,[x(-\tau),\rho_s(t)\right]]- \nu(\tau)\left[ y,[y(-\tau),\rho_s(t)\right]]- \notag \\
    &i \eta(\tau))\left[ y,[y(-\tau),\rho_s(t)\right]] \bigg \rbrace \label{mastereqBrownian}
\end{align}
where, $x(\tau)$ and $y(\tau)$ are the operators in the interaction picture and $\nu(\tau)$ and $\eta(t)$ are the noise and dissipation kernels respectively given by\cite{Schlosshauer2007,Schlosshauer2019quantum,Breuer2007}:
\begin{align}
   & \nu(\tau)=\int_0^\infty d\omega J(\omega)\coth\left( \frac{\omega}{\Omega_{th}}\right) \cos(\omega \tau) \label{noisekernel}\\
   & \eta(\tau)=\int_0^\infty d\omega J(\omega) \sin(\omega \tau) \label{dissipationkernel}
\end{align}
where, $\Omega_{th}=\frac{2 k_B T}{\hbar}$. \\ \\
$J(\omega)$ is the spectral density of the environment oscillators:
\begin{align}
    J(\omega)=\sum_i \frac{c_i}{2m_i \omega_i}\delta(\omega-\omega_i)
\end{align}
Now, we put the expressions for the operators $x(-\tau)$ and $y(-\tau)$ in Eq.(\ref{mastereqBrownian}) from Eqs.(\ref{solutionx}-\ref{solutiony}) and retain only the decoherence term in the time evolution of the reduced density matrix:
\begin{widetext}
\begin{align}
    &\frac{\partial \rho_s}{\partial t}=-\frac{1}{\hbar}\bigg[ \int_0^td\tau \nu(\tau)\frac{  \cosh(A \tau)+  \cosh(B\tau)}{2}\left[X,\left[X,\rho_s(t) \right]  \right]
     +\int_0^td\tau \nu(\tau)\frac{  \cosh(A \tau)+  \cosh(B \tau)}{2} \times \notag \\
    &\left[Y,\left[Y,\rho_s(t) \right]  \right] 
    +\alpha \int_0^td\tau \nu(\tau)f_0(-\tau)\left[X,\left[X^2,\rho_s(t) \right]  \right]+\alpha \int_0^td\tau \nu(\tau)f_1(-\tau)\left[X,\left[XY,\rho_s(t) \right]  \right]+  \notag \\
   & \alpha \int_0^td\tau \nu(\tau)f_2(-\tau)\left[X,\left[Y^2,\rho_s(t) \right]  \right]+\alpha \int_0^td\tau \nu(\tau)\cos(\omega_0 \tau)\left[Y,\left[Y^2,\rho_s(t) \right]  \right] \bigg] \label{decequation}
\end{align}
   \end{widetext}
   Notice that all the terms in Eq.(\ref{decequation}) contain Lindblad double commutators, pertaining to decoherence in terms. The first two terms represent the decoherence of the harmonic oscillator as seen in \cite{decoherence1,decoherence2}, whereas the last few $\alpha$-dependent terms originate from the anharmonicity in the system.\\
   The double commutator in the third term (anharmonicity-dependent) can be represented in the position basis as:
\begin{align}
   &\alpha\left[X,\left[X^2, \rho_s \right]\right]= \alpha\langle X',Y'|X^3 \rho_s -X \rho_s X^2 - \notag \\ 
   & X^2 \rho_s X + \rho_s X^3|X,Y\rangle\\
   &= \alpha\langle X',Y'| x'(x'^2-x^2)\rho_s - x(x'^2-x^2)\rho_s |X,Y\rangle \\
   &=\alpha(x'+x)(x'-x)^2 \rho_s(X,X',Y,Y',t)
\end{align}
This can be expressed in terms of the Wigner representation as $ 2 \alpha x \frac{\partial^2}{\partial p_x^2} W(x,y,p_x,p_y,t)$, where the Wigner function is given by : 
    \begin{align}
   & W(x,y,p_x,p_y,t)=\frac{1}{4\pi^2 \hbar^2}\int_{-\infty}^{\infty}du dv e^{\frac{i}{\hbar}(p_x u+p_y v)}  \notag\\
    &\rho(x+u/2,x-u/2, y+v/2, y-v/2,t) \label{wigner}
\end{align}
In a similar manner the last three terms in Eq.(\ref{decequation}) can also be written in the Wigner representation. The second derivatives in the Wigner representations of the alpha-dependent terms clearly signify the normal diffusive behaviour leading to decoherence. However, the diffusion coefficients are dependent on position coordinates in the presence of anharmonicity. \\
Thus, the terms retained in Eq.(\ref{decequation})  correspond to a loss of coherence and thus induce a quantum to classical transition, resulting in a loss of information from the system to the surroundings. 
The time evolution of the off-diagonal terms of the reduced density matrix calculated from Eq.(\ref{decequation}) is given by:
\begin{align}
    \rho_s(t)=\rho_s(0) \exp\left[-F_H(t)\right] \label{timerhored}
\end{align}
where, the heating function $F_H$ is given by:
\begin{align}
    F_H=\int_0^t h(t') dt' \label{timerho}
\end{align}
In Eq.(\ref{timerho}),
\begin{widetext}
\begin{align}
   & h(t)= -\frac{1}{\hbar}\bigg[ \int_0^td\tau \nu(\tau)\frac{  \cosh(A \tau)+  \cosh(B\tau)}{2}(\Delta_x)^2
     +\int_0^td\tau \nu(\tau)\frac{  \cosh(A \tau)+  \cosh(B \tau)}{2} (\Delta_y)^2
    +\notag \\
    &2\alpha \int_0^td\tau \nu(\tau)f_0(-\tau) \Bar{x}(\Delta_x)^2+\alpha \int_0^td\tau \nu(\tau)f_1(-\tau) \Delta_{xy} \Delta_x+   2\alpha \int_0^td\tau \nu(\tau)f_2(-\tau)  \Bar{y} \Delta_x \Delta_y+\notag \\
   &2\alpha \int_0^td\tau \nu(\tau)\cos(\omega_0 \tau) \Bar{y}(\Delta_y)^2 \bigg]
\end{align}
\end{widetext}
where, $\Delta_x= (x'-x)$, $\Delta_y=(y'-y)$, \\
$\Delta_{xy}=(x'y'-xy)$, $\Bar{x}=(x'+x)$, $\Bar{y}=(y'+y)$\\
Thus, Eq. (\ref{timerhored}) displays the temporal suppression of the off-diagonal elements of the reduced density matrix within the framework of the non-Markovian dyanamics \cite{Schlosshauer2019quantum}.  This decay of the reduced density matrix $\rho_s(t)/\rho_s(0)$ captures decoherence of the system, leading to the loss of information.
\section{von Neumann Entropy}
The von Neumann entropy which quantifies the information contained in the system can be obtained from the reduced density matrix via the following relation\cite{wheeler,Horhammer2008}:
\begin{align}
    S=-k_B\mathrm{Tr} (\rho_s \ln \rho_s) \label{Vonentropy}
\end{align}
So, the von Neumann entropy in Eq.(\ref{Vonentropy}) can be expressed as $S=-k_B\langle \ln\rho_s \rangle$, only if the off diagonal terms of the density matrix are zero in the position basis representation, which is not the case in the low temperature quantum regime \cite{weiderpass2020}. However, in the high temperature regime, one can compute the von Neumann entropy by calculating the Wigner function and its Weyl transformed form \cite{weiderpass2020}. \\ 
The density operator is related to the Wigner function as follows\cite{Isar1999}:
 \begin{align}
\rho_s=4 \pi^2 \hbar^2 \mathcal{N}\lbrace W_s(\mathbf{x,y,p_x,p_y})\rbrace \label{wignerdensityrelation}
\end{align}
where $W_s$ is the Wigner function in the form of the standard rule of association and $\mathcal{N}$ is the normal ordering operator. This relation sets a correspondence between variables in classical mechanics and operators in quantum mechanics. Note that the Wigner function displayed here is considered in the steady state asymptotic limit of $t\rightarrow\infty$ and therefore does not have any explicit time dependence \cite{Isar1999}.
The Weyl transform in two dimensions, which serves as the basis for the normal ordering between $p$ (momentum) and $q$  (position) in the Wigner function is given by (see Appendix-B):
 \begin{align}
    &W_s\mathbf{(x,y,p_x,p_y)}=  \notag \\
    &\exp \bigg[ \frac{i\hbar} {2}\left( \frac{\partial^2}{\partial p_x \partial x} + \frac{\partial^2}{\partial p_y \partial y} \right) \bigg] W(x,y,p_x,p_y) \label{Weyl}
\end{align}
 Here $\mathbf{x,y,p_x}$ and $\mathbf{p_y}$ on the L.H.S. of Eq.(\ref{Weyl}) are operators and obey non-trivial commutation relations. Thus the Weyl transform leads to a transition from classical observables to operators in quantum mechanics. So boldface is used in Eqs.(\ref{wignerdensityrelation}) and (\ref{Weyl}) to clearly differentiate between the classical observables and quantum mechanical operators. In all others equations, we deal with operators and so for simplicity we have used the notations $x,y,p_x,p_y$ (dropping the boldface) for the operators throughout the rest of the paper. \\
For an anharmonic oscillator in the presence of a magnetic field, the Wigner function (Eq.(\ref{wigner})) in the standard rule of association is as follows \cite{Isar1999,dickman1986}:
\begin{widetext}
\begin{align}
    &W_s(x,y,p_x,p_y)= \left[\frac{1}{4\pi^2 \eta^2} \exp \left[ -\frac{1}{ 
    \eta \omega_0} \left \lbrace V_{HM}+\frac{(p_x+\frac{eBy}{2})^2}{2m }+ \frac{(p_y-\frac{eBx}{2})^2}{2m}-m\omega_0^2 \alpha x^3\right \rbrace  \right] \right]_s \label{Wigner}\\
    &W_s(x,y,p_x,p_y)= \left[\frac{1}{4\pi^2 \eta^2} \exp \left[ -\frac{1}{ 
    \eta \omega_0} \left \lbrace H_0- \alpha H'\right \rbrace  \right] \right]_s \label{perturbationwigner}\\
    \text{where,} \\
    &H_0=V_{HM}+\frac{(p_x+\frac{eBy}{2})^2}{2m }+ \frac{(p_y-\frac{eBx}{2})^2}{2m} \label{unperturbedHamiltonian}\\
    &H'=m\omega_0^2 x^3 \label{perturbedHamiltonian}
\end{align}
\end{widetext}
$V_{HM}$ in the unperturbed system Hamiltonian $H_0$(Eq.(\ref{unperturbedHamiltonian}))  corresponds to the harmonic oscillator potential given by:
\begin{align}
V_{HM}=\frac{1}{2}m \omega_0^2(x^2+y^2)
\end{align}
$ \eta$ is the determinant of the dispersion (correlation) matrix, the asymptotic value of which depends on the diffusion coefficients \cite{Isar1999}. The anharmonic potential in  $H'$ (Eq.(\ref{perturbedHamiltonian})) is treated as a perturbation to $H_0$ as discussed in the preceding sections. The applied magnetic field $B$ is in the $z$-direction, originating from the vector potential $A$ in the $x-y$ plane ($\vec{B}=\vec{\nabla} \times \vec{A}$).  \\
Here, we derive the normal ordered Wigner function applying Eq.(\ref{Weyl}) and Eq.(\ref{Wigner}) and use  $S=-k_B\langle \ln\rho_s \rangle$ (see Eq.(\ref{Vonentropy})) to compute the von Neumann entropy of the anharmonic oscillator (see Appendix-C):\\
\begin{align}
  S_{VN}&= -k_B Tr (\rho \ln \rho) \notag \\
  &=(S_{VN})_{HM} -k_B \sum_j \langle j| (\rho \ln\rho)_{A}  |j \rangle \label{entropy1}
\end{align}
where, $(\rho \ln\rho)_A$ and $(S_{VN})_{HM}$ in Eq.(\ref{entropy1}) denote the anharmonic part of $(\rho \ln\rho)$ and the $\alpha$-independent harmonic part of the von Neumann entropy respectively.\\
Then the von Neumann entropy is calculated considering the magnetic field to be weak and the anharmonicity treated as a small perturbation on the harmonic oscillator Hamiltonian \cite{Robinnet19977}. In the derivation of Eq.(\ref{entropy1}) from Eq.(\ref{Wigner}), the terms containing $\alpha$ vanish. So one retains the $\alpha^2$ term as the lowest order perturbative term affecting the von Neumann entropy of the anharmonic oscillator (Eq.(\ref{entropy1})).
Thus we obtain the total von Neumann entropy (see Appendix-C):
\begin{align}
    &S_{VN}=(S_{VN})_{HM}+ \nonumber\\
    &\frac{9 k_B\hbar^6 \alpha^2}{32 m \omega_0 \eta^5}\left[(2 n_x+1)^2+2(n_x^2+n_x+1) \right]
    \label{Vonneumann}
\end{align}
Eq.(\ref{Vonneumann}) represent the von-Neumann entropy of the anharmonic oscillator in the presence of a weak magnetic field and coupled to an external heat bath. The von Neumann entropy pertaining to the harmonic oscillator ($(S_{VN})_{HM}$)  can be recovered by setting the anharmonicity parameter $\alpha=0$ in Eq. (\ref{Vonneumann}). Here, $n_x$ is the expectation value of the number operator of harmonic oscillator states in the long time limit \cite{Agarwal1971}. Thus, from the above equation one can notice that the anharmonicity results in an increase in the von Neumann entropy and correspondingly a stronger deconfinement of the system Brownian particle is observed for higher values of the anharmonicity parameter $\alpha$.
\section{ Proposed experimental setup }
\begin{widetext}
  
\begin{figure}[H]
\centering
\fbox{\includegraphics[scale=0.6]{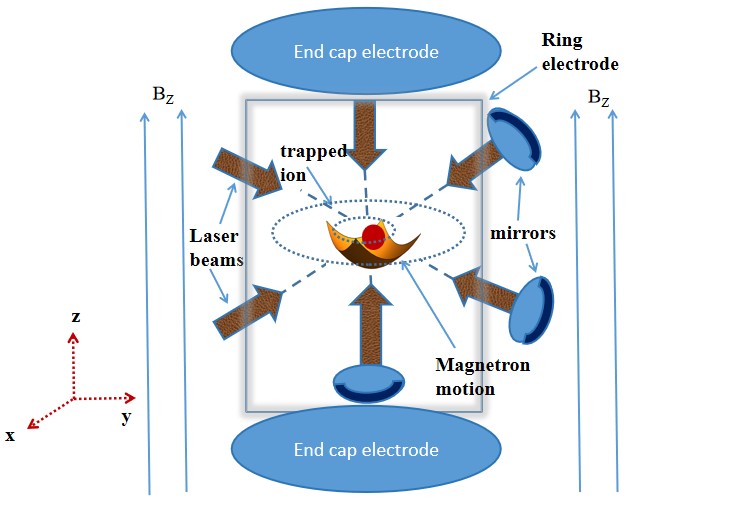}} 
\caption{Proposed experimental setup for an anharmonically oscillating charged particle trapped in a Penning trap. The counter-propagating laser beams are used to cool the trapped ions and act as an optical molasses. There is an 
externally applied magnetic field in the $z$-direction as shown in the figure.} 
\label{expt}
\end{figure}

\end{widetext}

 An experimental setup can be designed to test the predictions of our theoretical analysis. 
 Hybrid traps for ions and neutral atoms can be used for confining a charged cloud \cite{Tomza2019,Cui2025}. In this paper, we propose a Penning trap set up, which can trap charged ions in the presence of electric and magnetic fields \cite{Blaum2020}. The electric field is produced by ring electrodes and end cap electrodes at the two ends of the set up as shown in Fig.(\ref{expt}). The counter-propagating circularly polarized laser beams cool the trapped ions via a laser cooling technique involving a magnetic field, as done in a MOT (magneto-optical trap) \cite{Misra2025,Bhar2022}. 
An optical molasses is created by the laser beams and it 
mimics the Ohmic heat bath in our theoretical model. 
The charged cloud in an anharmonic potential in the presence of an applied magnetic field
undergoes decoherence as a result of its coupling to the optical molasses bath. Our proposed experimental set up using a Penning ion trap and the laser beams as optical molasses is shown in Fig.(\ref{expt}).  Such a cold ion setup can then be used to make measurements related to the decoherence effects predicted here. For instance, one can manipulate the motional states of the ion using laser pulses and the quantum dephasing can be measured. 
 In order to test our theoretical predictions, the parameters in the experiment should be as follows: 
one needs to consider the trap frequency $\omega_0/2\pi$ in the $MHz$ range and the cyclotron frequency $\omega_c/2\pi$ in the 
$kHz$ range. The maximum value of the anharmonicity parameter $\alpha$ needs to be of the order of $0.1 m^{-1}$ for the anharmonicity to be in 
the perturbative regime for our theoretical analysis to be applicable. We have predictions both for the high and low temperature regimes. Thus the bath temperature needs to be respectively in the $mK$ and $\mu K$ range in the high and low temperature regimes. \\
Further, the Ramsey interferometry can be used to measure the decoherence time in a Penning trap and the resolution of this method 
can be improved by a suitable choice of the Rabi frequency and the detuning \cite{Wan2016}.\\
While it is complicated to design an experiment for measuring the density matrix and entanglement in an open quantum system, however, some cold atom simulators have been developed to mimic the experimental set up, which can give an insight into the entanglement and the von Neumann entropy \cite{Zoller2018,mendes2020}. Moreover, in \cite{huang2023}, the authors have used a Mach–Zehnder (M-Z) interferometer to create mixed state of qubits with photonic systems. The von Neumann entropy is measured for the system, leading to the direct measurement of relative entropy of coherence \cite{Nikola2022,de2020proposal,huang2023}.
%

\section{ Results and Discussions}  
In Fig.(\ref{decoherence}) and Fig.(\ref{heating function}), we have plotted the time variation of the off-diagonal elements of the reduced density matrix ($\rho_s/\rho_s(0)$) and the function $h(t)$ respectively in the low and high temperature regimes. The plots have been displayed for different values of the anharmonicity parameter $\alpha$.  Fig.(\ref{heating function non-Markovian}) displays the heating functions $F_H$ versus time for the non-Markovian and Markovian cases respectively. The plots in Fig.(\ref{Von Neumann}) show the scaled von Neumann entropies of the anharmonicity dependent part, corresponding to different harmonic oscillator states $n_x$ and different harmonic frequencies $\omega_0$.
\begin{widetext}

\begin{figure}[H]
\centering
\hspace*{-1cm}\includegraphics[scale=1.32]{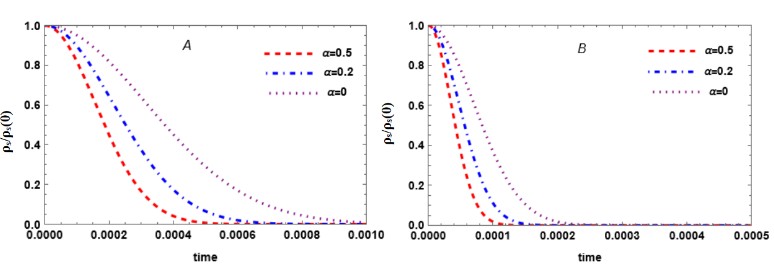}
\caption{Reduced density matrix $\rho_s/\rho_s(0)$ as a function of time for different values of the anharmonicity parameter $\alpha$ with $\omega_0=10$, $\omega_c=0.1$, $\gamma=10$, $x'=2$, $x=1$, $\Lambda=10^3$: (A) Low temperature ($\Omega=0.1$), (B) High temperature ($\Omega=10^4$).} 
\label{decoherence}
\end{figure}

\begin{figure}[H]
\centering
\hspace*{-1.4cm}\includegraphics[scale=1.19]{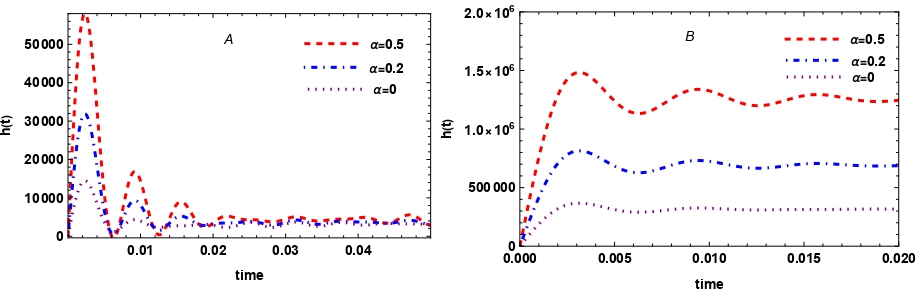} 
\caption{h(t) versus time for different values of the anharmonicity parameter $\alpha$ with $\omega_0=10$, $\omega_c=0.1$, $\gamma=10$, $x'=2$, $x=1$, $\Lambda=10^3$: (A) Low temperature ($\Omega=0.1$), (B) High temperature ($\Omega=10^4$).} 
\label{heating function}
\end{figure}
\begin{figure}[H]
\centering
\hspace*{-1cm}\includegraphics[scale=0.675]{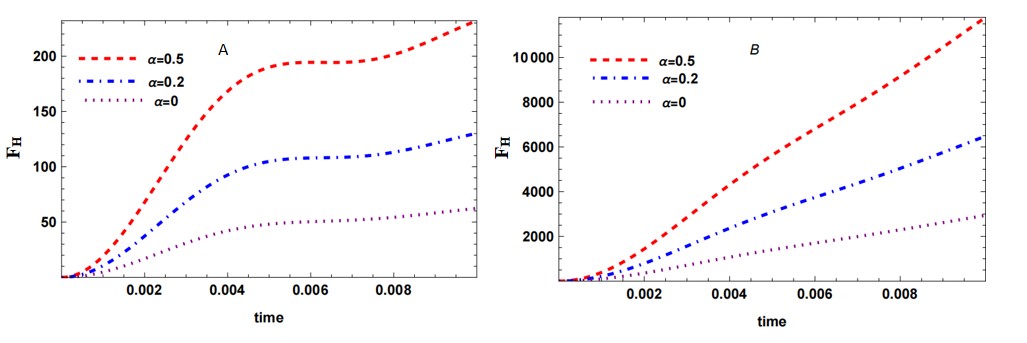}
\hspace*{-1cm}\includegraphics[scale=0.69]{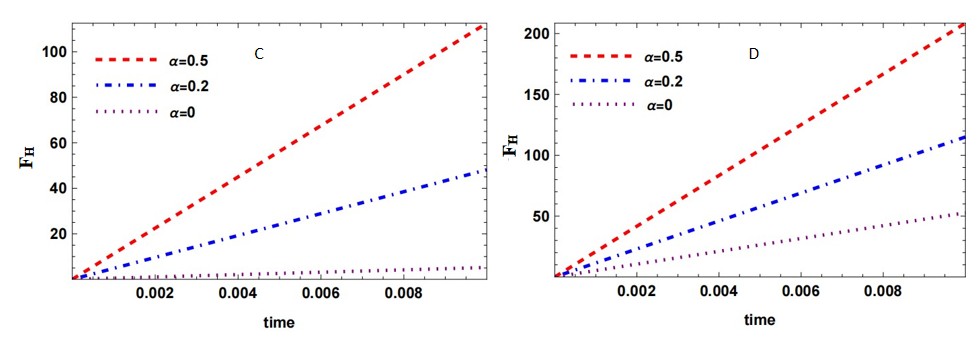} 
 \caption{Heating function $F_H$ versus time for different values of the anharmonicity parameter $\alpha$ with $\omega_0=10$, $\omega_c=0.1$, $\gamma=10$, $x'=2$, $x=1$, $\Lambda=10^3$ for non-Markovian cases: (A) Low temperature ($\Omega=0.1$), (B) High temperature ($\Omega=10^4$); and Markovian cases:  (C) Low temperature ($\Omega=0.1$), (D) High temperature ($\Omega=10^4$) } 
\label{heating function non-Markovian}
\end{figure}

\begin{figure}[H]
\centering
\hspace*{-1cm}\includegraphics[scale=1.04]{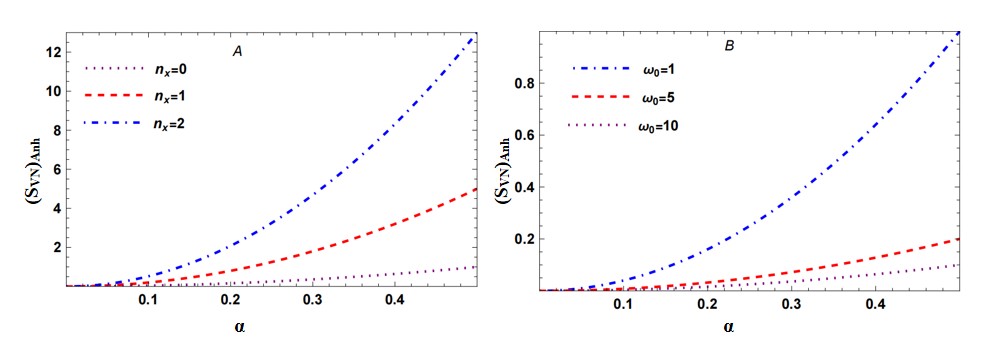} 
\caption{Anharmonic part of the von Neumann entropy ($(S_{VN})_{Anh}=S_{VN}(\alpha,n_x)/S_{VN}(1/2,1)$)} as a function of $\alpha$ for different values of (A) $n_x$ and (B) harmonic frequency $\omega_0$.
\label{Von Neumann}
\end{figure}

\end{widetext}
\vspace*{-1.6cm}
 \subsection{Decoherence and heating function }
In Fig.(\ref{decoherence}), we notice that the rate of decay of the off-diagonal terms of the reduced density matrix ($\rho_s/\rho_s(0)$) increases with the increase in the anharmonicity. Thus one expects a faster loss of information from the system to the surroundings with an increase in the anharmonicity, resulting in a quantum to classical transition. Here, the plots have been presented in the low cyclotron frequency limit in order to highlight the effect of the applied harmonic potential and the anharmonic perturbation on the system.  The increased anharmonicity results in a greater deconfinement of the particle, and thus a faster 
 quantum dephasing is observed for higher values of $\alpha$. This loss of information from the system is even faster in the high temperature regime (Fig.(\ref{decoherence}B)) compared to the low temperature quantum-fluctuation dominated regime (Fig.(\ref{decoherence}A)), 
as had been reported in several earlier works \cite{decoherence1,decoherence2}. In Fig.(\ref{heating function}), we have plotted the function $h(t)$ against time in the low (Fig.(\ref{heating function}A)) and high (Fig.(\ref{heating function}B)) temperature regimes. The plots show oscillations that gradually die out at longer times  and it settles down to its time-independent Markovian value in the long time limit. This is similar to the oscillations of the heating function (which is the time integral of $h(t)$) observed in some of the earlier works \cite{Malay2015,malay2,Paavola}. One notices that these oscillations are more prominent in the low temperature regime than in the high temperature classical domain, where the thermal fluctuations lead to a faster decoherence of the system. 
 Moreover, in Fig.(\ref{heating function non-Markovian}) we have plotted the heating function for both Markovian and non-Markovian cases at low and high temperatures. For the Markovian case, the function $h(t)$ is replaced by its value in the Markovian limit, where the observation time is much larger than the bath relaxation time. The heating function $F_H$ signifies the heating of the system due to absorption of energy from the environment \cite{Malay2015}. As in the plot of $h(t)$ versus time, here too we notice non-monotonicity in the plots of the non-Markovian case, stemming from the backflow of lost information from non-trivial correlations in the bath. On the other hand, bath correlations decay fast in the Markovian case and non-monotonicity is not there  due to the absence of memory effects. The heating functions in the high temperature regimes display a faster rise with time in both Markovian and non-Markovian cases as the heat absorption from the environment and the loss of coherence are faster 
as intuitively expected. In addition, one notices that the rise in $F_H$ is faster for the anharmonic oscillator model, compared to the heating functions observed in the earlier works based on a harmonic oscillator model \cite{decoherence1,decoherence2,Malay2015}. We also notice from Fig.(\ref{heating function non-Markovian}) that the heating function is higher for larger values of the anharmonicity parameter $\alpha$, as the heat absorption from the environment is enhanced due to the deconfining effect anharmonicity in the system.\\
 \subsection{von Neumann entropy : an information theoretic measure}
The von Neumann entropy is calculated for the anharmonic oscillator system to obtain a measure of the system-environment entanglement. We apply the Weyl theory (Eq.(\ref{Weyl},\ref{Wigner})) at high temperatures to derive the normal-ordered Wigner function (see Appendix-B and C) \cite{folland2016,curtright2013,wheeler}. The Wigner function, representing the quasi-probability distribution in phase space turns out to be positive, when calculated in the Gaussian states of a harmonic oscillator model \cite{soto1983,curtright2001,nedjalkov2011}. The function can take negative values in certain regions of phase space in the case of non-Gaussian states, however, it has been shown that positive Wigner functions are also plausible for non-Gaussian states, including anharmonic oscillators \cite{corney2015,filip2011}. Here, a small anharmonicity is considered as a perturbation on the harmonic hamiltonian and truncated to get a positive Wigner function that corresponds to a classical probability distribution through the Weyl correspondence formula for two dimensional systems. The formalism is used to calculate the von-Neumann entropy (Eq.(\ref{entropy1})) pertaining to the anharmonic oscillator coupled to a bath of harmonically oscillating tiny particles. The entropy calculation shows that the $\alpha$-dependent terms in $\ln(\rho_s)$ go to zero, when the expectation values are taken in the harmonic oscillator states. So $\alpha^2$ is retained as the lowest order of anharmonicity affecting the von Neumann entropy of the system. We notice in Eq.(\ref{entropy1}) that the von-Neumann entropy increases with the anharmonicity, resulting in a greater deconfinement in the system with the increase in anharmonicity.  In Figs.(\ref{Von Neumann}A,B), we have plotted the anharmonic part of the von Neumann entropy ($S_{VN}(\alpha,n_x)$), scaled by the entropy at $\alpha=1/2$ and $n_x=1$ ($S_{VN}(1/2,1)$). The plots are displayed as functions of $\alpha$, corresponding to various $n_x$ and $\omega_0$. From the figures, one notes that the anharmonicity results in an increase 
in the entropy in all the displayed cases and the enhancement is more pronounced for higher values of $n_x$ (Fig.(\ref{Von Neumann}A)) and lower values of the harmonic frequency $\omega_0$ (Fig.(\ref{Von Neumann}B)). The higher values of the von Neumann entropy ($S_{VN}$) signify a faster loss of information stemming from the system-bath entanglement and the statistical uncertainty associated with the mixed states \cite{Pucci2013}. $S_{VN}$ goes to zero in the case of pure states and serves as a measure of the extent of deviation from purity of quantum states, even at zero temperature. So the von Neumann entropy is a measure of the accessible quantum information accessible, expressed in terms of the reduced density matrix $\rho_s$ of the subsystem \cite{Pucci2013,Horhammer2008}. In contrast, the well known thermodynamic entropy deals only with the amount of heat contained in the system and is central to the establishment of the thermodynamic laws \cite{Horhammer2008,scripta2010}. Thus, it turns out that the thermodynamic entropy is smaller than the von Neumann entropy as shown in \cite{Horhammer2008,scripta2010}. In this paper we have confined our attention to the analysis of the von Neumann entropy as it is the relevant entropy for exploring quantum entanglement and information loss in an open quantum system and that is the central focus of our study where we view decoherence from the perspective of quantum information \cite{Isar1999,Pucci2013}. The faster rise in the von Neumann entropy with $\alpha$ for lower values of $\omega_0$ is attributed to the lower dominance of harmonic confinement, leading to a stronger effect of anharmonicity on the particle.  Thus, it can be inferred that the effect of anharmonicity on the von Neumann entropy is comparatively less (of order $\alpha^2$), however, it clearly shows an increase in deconfinement in the system, in agreement with the results obtained from our decoherence calculation. A similar effect can be seen for an anharmonic oscillator coupled to a Bose Einstein condensate trapped in a symmetric double well potential
\cite{Alonso2014}. The authors have shown that an increase in the anharmonicity induces faster decoherence and non-exponential anomalous 
quantum dephasing is noted in different time regimes, as can also be seen from our theoretical and numerical results.  
\subsection{Experimental implications}
Several experimental techniques have also been developed to control the decoherence 
in open quantum systems, including the decay of the mesoscopic superposition of quantum states \cite{Brune1996}. Furthermore, anomalous diffusion was studied in ultracold Rydberg atoms in optical lattices \cite{cardman2020,aman2016,raitzsch2009}. In recent times, many experiments use lattice traps, built by reflecting laser beams off the membranes of optical mirrors \cite{ian2008,camerer2011}. Moreover, charged particles can be trapped using hybrid traps that trap neutral particles using a MOT set up and ions with RF traps or Penning traps, involving electric and magnetic fields \cite{Peik1999,Bhar2022,Misra2025,Blaum2020,Gabrielse2009}.  The experiments also demonstrate that decoherence in the systems can be tuned by coherently manipulating the states of the harmonic oscillator \cite{meiser2006,genes2008}. In Sec. $5$ (Fig.(\ref{expt})), we have proposed a suitable ion trap in the presence of a magnetic field and an anharmonic potential that can be used to test our theoretical predictions regarding the dynamics and decoherence of a charged particle in the presence of a weak magnetic field and coupled to an Ohmic heat bath.
\section*{Acknowledgement}{We thank Sanjukta Roy for useful discussions regarding the experimental realisation of our predictions.}
\begin{widetext}
\section*{Appendix}
\subsection*{A: Perturbative solution}
\renewcommand{\theequation}{A.\arabic{equation}}
\setcounter{equation}{0}

We solve the coupled differential equation in Eq. (\ref{perturb2}) to get $x_1$ and $y_1$ in terms of the initial values of the x and y coordinates:
\begin{align}
 X\Ddot{x_1}(t) + X\omega_0^{2} x_1(t) -3\omega_0^{2} x_0^2(t)-Y\omega_{c}\dot{y_1}(t) =0 \notag\\
    Y\Ddot{y_1}(t)+ Y\omega_0^{2} y_1(t) +X\omega_{c}\dot{x_1}(t)= 0 \notag
\end{align}
Solving these equations in the small $\omega_c$ limit we get:
\begin{align}
   & x_{1}(t)= f_{0} (t) X + f_{1} (t) Y +  \notag \\
    &f_{2} (t) (Y^{2}/X) +\frac{\sin(\omega_0 t)}{\omega_0} V_x \label{x1sol1}\\
     &y_{1}(t)= Y \cos(\omega_0  t) +\frac{V_y}{\omega_0} \sin(\omega_0 t) \label{y1sol1}
    \end{align}
where the functions $f_0$, $f_1$ and $f_2$ in the small $\omega_c$ limit are given by:
\begin{align}
f_0(t)&=C_0-C_1 \cos \left(\omega_0 t \right)-C_2 \cos \left(2 A t\right)+C_3 \cos\left(A t \right) \cos\left(B t \right)+ C_4 \cos\left(2B t \right)- \notag \\
&C_5 \sin \left( A t\right) \sin \left( B t\right)
   \\ \notag \\
    f_1(t)&=C_6 \sin\left( \omega_0 t \right)+i C_7 \cos\left(B t \right) \sin \left( A t\right)-i C_8 \sin\left(2A t \right)+C_9 \cos\left( A t\right)+ \notag \\
    &i C_{10} \cos \left(B t \right) \sin \left(B t \right)\\ \notag \\
     f_2(t)&=C_{11}+C_{12} \cos\left(\omega_0 t\right)+C_{13}\cos \left( 2A t\right)- C_{14} \cos \left(2B t \right)+ 
    C_{15}\cos\left(A t\right)\cos\left(B t\right)-\notag \\
    &C_{16} \sin\left(A t\right)\sin\left(B t\right)\\ \notag \\
    &A=\sqrt{\omega_0(\omega_0+\omega_c)} \hspace*{0.2cm}, \hspace*{0.4cm}B=\sqrt{\omega_0(\omega_0-\omega_c)}
\end{align}
Here, the coefficients of the sine and cosine terms ($C_0-C_{16}$) are constants dependent on the parameters $\omega_0$ and $\omega_c$.
\subsection*{B: Weyl transform}
\renewcommand{\theequation}{B.\arabic{equation}}
\setcounter{equation}{0}    
The classical observable $a(\alpha,\beta,\gamma,\delta)$ can be transformed into a quantum mechanical operator  $\mathbf{A(x,y,p_x,p_y)}$ through an operator-based Fourier transform (all operators are indicated in boldface) \cite{wheeler}:
\begin{align}
     \mathbf{A(x,y,p_x,p_y)}=\int \int \int \int a(\alpha, \beta, \gamma, \delta) e^{\frac{i}{\hbar}(\alpha \mathbf{p_x}+\beta \mathbf{x}+\gamma \mathbf{p_y}+\delta \mathbf{y}}) d \alpha d \beta d \gamma d \delta \label{2DWeyl}
\end{align}
Let us define the operator $\mathbf{E}(\alpha,\beta,\gamma,\delta)$ as follows\cite{wheeler}:
\begin{align}
    mathbf{E}(\alpha,\beta,\gamma,\delta)=e^{\frac{i}{\hbar}\lbrace(\alpha \mathbf{p_x}+\gamma \mathbf{p_y})+(\beta \mathbf{x}+\delta \mathbf{y})\rbrace} \label{Ealphabeta}
\end{align}
Now, considering up to the second order in the Baker-Campbell-Hausdroff formula where the operators  \textbf{X} and  \textbf{Y} do not commute \cite{bonfiglioli2011} we get:
 \begin{align}
    &e^{\mathbf{X}}e^{\mathbf{Y}}=e^{\mathbf{Z}}\\
    &\mathbf{Z}=\mathbf{X}+\mathbf{Y}+\frac{1}{2}\left[\mathbf{X},\mathbf{Y}\right] \label{BakerHausdroff}
\end{align}
So, applying Eq.(\ref{BakerHausdroff}) in Eq.(\ref{Ealphabeta}) we get:
 \begin{align}
   e^{\frac{i}{\hbar}(\alpha \mathbf{p_x}+\gamma \mathbf{p_y})} .e^{\frac{i}{\hbar}(\beta \mathbf{x}+\delta \mathbf{y})} &=\exp\left[\frac{i}{\hbar}(\alpha \mathbf{p_x}+\gamma \mathbf{p_y}+\beta \mathbf{x}+\delta \mathbf{y})+\frac{1}{2}\left[\frac{i}{\hbar}(\alpha \mathbf{p_x}+\gamma \mathbf{p_y}),\frac{i}{\hbar}(\beta \mathbf{x}+\delta \mathbf{y}) \right] \right ] \\
   &=e^{\frac{i}{\hbar}(\alpha \mathbf{p_x}+\gamma \mathbf{p_y}+\beta \mathbf{x}+\delta \mathbf{y})}e^{\frac{i}{2 \hbar}(\alpha \beta+\gamma \delta)} \label{pordered1} 
   \end{align}
   So from Eq.(\ref{pordered1}) we have,
   \begin{align}
   &e^{\frac{i}{\hbar}(\alpha \mathbf{p_x}+\gamma \mathbf{p_y}+\beta \mathbf{x}+\delta \mathbf{y})}=e^{\frac{i}{\hbar}(\alpha \mathbf{p_x}+\gamma \mathbf{p_y})} .e^{\frac{i}{\hbar}(\beta \mathbf{x}+\delta \mathbf{y})}e^{-\frac{i}{2 \hbar}(\alpha \beta+\gamma \delta)}
   \label{pordered2}
\end{align}
Eq.(\ref{pordered2}) represents the $\mathbf{p-x}$ ordered form. Similarly, the $\mathbf{x-p}$ ordered form will give:
\begin{align}
 &e^{\frac{i}{\hbar}(\beta \mathbf{x}+\delta \mathbf{y})}.e^{\frac{i}{\hbar}(\alpha \mathbf{p_x}+\gamma \mathbf{p_y})}e^{\frac{i}{2 \hbar}(\alpha \beta+\gamma \delta)}=e^{\frac{i}{\hbar}(\beta \mathbf{x}   +\delta \mathbf{y} +\alpha \mathbf{p_x} +\gamma \mathbf{p_y} )} \label{xordered}
 \end{align}
 So the operator $ \mathbf{E}(\alpha,\beta,\gamma,\delta)$ takes the form:
 \begin{align}
 \mathbf{E}(\alpha,\beta,\gamma,\delta)=e^{\frac{i}{\hbar}(\beta \mathbf{x}+\delta \mathbf{y}+\alpha \mathbf{p_x}+\gamma \mathbf{p_y})}=e^{\frac{i}{2 \hbar}(\alpha \beta+\gamma \delta)}e^{\frac{i}{ \hbar}(\beta \mathbf{x}+\delta \mathbf{y})}e^{\frac{i}{ \hbar}(\alpha \mathbf{p_x}+\gamma \mathbf{p_y})}
\end{align}
We apply Eq.(\ref{xordered}) in Eq.(\ref{2DWeyl}) to derive the operator \textbf{A} from the classical observable:
 \begin{align}
    \textbf{A}=&\int\int \int \int a(\alpha,\beta,\gamma,\delta) e^{\frac{i}{2 \hbar}(\alpha \beta+\gamma \delta)}e^{\frac{i}{ \hbar}(\beta \mathbf{x}+\delta \mathbf{y})}e^{\frac{i}{ \hbar}(\alpha \mathbf{p_x}+\gamma \mathbf{p_y})} d \alpha d \beta d \gamma d \delta \label{Weylcal1}    \\
    &=\bigg{.} ^{}_{\mathbf{xy}}\left[\int\int \int \int a(\alpha,\beta,\gamma,\delta) e^{\frac{i}{2 \hbar}(\alpha \beta+\gamma \delta)}e^{\frac{i}{ \hbar}(\beta x+\delta y)}e^{\frac{i}{ \hbar}(\alpha p_x+\gamma p_y)} d \alpha d \beta d \gamma d \delta \right]_{\mathbf{p_xp_y}} \\
    &={{}_{\mathbf{xy}}[A_{xyp_xp_y}(x,y,p_x,p_y)]_{\mathbf{p_xp_y}}} \label{Weylcalc}
\end{align}
where, ${^{}_{\mathbf{xy}}[A_{xyp_xp_y}(x,y,p_x,p_y)]_{\mathbf{p_xp_y}}}$ represents the operator based Fourier transform (as shown in Eq.(\ref{2DWeyl})) of the function $A_{xyp_xp_y}(x,y,p_x,p_y)$,
which shows the elegance of Weyl transform in establishing the correspondence between a quantum mechanical operator and a classical observable. 
In Eq.(\ref{Weylcal1}), the x-p ordered form of the function $A_{xyp_xp_y}(x,y,p_x,p_y)$ is given by:
\begin{align}
A_{xyp_xp_y}(x,y,p_x,p_y)=e^{\frac{i}{2 \hbar}(\alpha \beta+\gamma \delta)}A(x,y,p_x,p_y)=\exp\left[\frac{i\hbar}{2}\left(\frac{\partial^2}{\partial p_x \partial x} +\frac{\partial^2}{\partial p_y \partial y}\right) \right] A(x,y,p_x,p_y)\label{Weylcalcbthrtn}
\end{align}
with  
\begin{align}
A(x,y,p_x,p_y) =\int\int\int\int a(\alpha,\beta,\gamma,\delta) e^{\frac{i}{ \hbar}(\beta x+\delta y)}e^{\frac{i}{ \hbar}(\alpha p_x+\gamma p_y)} d \alpha d \beta d \gamma d \delta
\end{align}
Eq. (\ref{Weylcalcbthrtn}) can be easily checked by an expansion of the exponential factor on the right hand side.
\subsection*{C: von-Neumann entropy}
\renewcommand{\theequation}{C.\arabic{equation}}
\setcounter{equation}{0}
We expand the Weyl transform defined in Eq.(\ref{Weyl}) as follows:
\begin{align}
    W_s(x,y,p_x,p_y)=&\bigg[1+\frac{i\hbar}{2}\partial^2_{p_x,x}+\frac{i\hbar}{2}\partial^2_{p_y,y}- \frac{1}{2!}\frac{\hbar^2}{4}  \partial^4_{p_x,x} - \frac{1}{2!}\frac{\hbar^2}{4}  \partial^4_{p_y,y}- \notag \\
    &\frac{1}{2!}\frac{\hbar^2}{4}  \partial^4_{p_y,y,p_x,x}-\frac{1}{2!}\frac{\hbar^2}{4}  \partial^4_{p_x,x,p_y,y}\bigg]W(x,y,p_x,p_y) \label{Weylexpansion}
    \end{align}
    where,
    \begin{align*}
        &\partial^2_{p_x,x}=\frac{\partial^2}{\partial_{p_x}\partial_x} \hspace*{0.2cm}, \hspace*{0.3cm} \partial^2_{p_y,y}=\frac{\partial^2}{\partial_{p_y}\partial_y}, \hspace*{0.3cm}\partial^4_{p_x,x}=\frac{\partial^2}{\partial_{p_x}\partial_x}\left(\frac{\partial^2}{\partial_{p_x}\partial_x}\right), \hspace*{0.3cm} \partial^4_{p_y,y}=\frac{\partial^2}{\partial_{p_y}\partial_y}\left(\frac{\partial^2}{\partial_{p_y}\partial_y}\right)\\
        &\partial^4_{p_y,y,p_x,x}=\frac{\partial^2}{\partial_{p_y}\partial_y}\left(\frac{\partial^2}{\partial_{p_x}\partial_x}\right) \hspace*{0.3cm}\partial^4_{p_x,x,p_y,y}=\frac{\partial^2}{\partial_{p_x}\partial_x}\left(\frac{\partial^2}{\partial_{p_y}\partial_y}\right)
    \end{align*}
    Now expanding the terms in Eq.(\ref{Weylexpansion}), one gets:
    \begin{align}
& \textbf{(i)} \frac{i\hbar}{2}\partial^2_{p_x,x}= \frac{i \hbar\left(2p_x+m\omega_c  y\right)\left(-2 \omega_c p_y +m \omega_c^2  x+4 m x(1+3 x \alpha)\omega_0^2\right)}{64 m \pi^2 \eta^4 \omega_0^2}  \exp \left[-\frac{1}{ \eta \omega_0}\lbrace H_0 -\alpha H'\rbrace \right] \\
&\textbf{(ii)} \frac{i\hbar}{2}\partial^2_{p_y,y}= \frac{i\hbar\left(2p_y-m\omega_c  x\right)\left(2 \omega_c p_x +m \omega_c^2  y+4 m y\omega_0^2\right)}{64 m \pi^2 \eta^4 \omega_0^2}  \exp \left[-\frac{1}{ \eta \omega_0}\lbrace H_0 -\alpha H'\rbrace \right] \\
&\textbf{(iii)} -\frac{1}{2}\frac{\hbar^2}{4}\partial^4_{p_x,x}=-\frac{\hbar^2}{2048 m^2 \pi^2 \eta^6 \omega_0^4}\bigg[((2p_x+m\omega_c  y)^2-4 m \eta \omega_0) (\omega_c^2(-2 p_y+m\omega_c x)^2-4 m \omega_c^2\eta \omega_0+\notag \\
&8 m \omega_c x(-2p_y+m \omega_c x)(1+3 x \alpha)\omega_0^2-16m(1+6 x \alpha)\eta \omega_0^3+16m^2 x^2(1+3 x \alpha)^2 \omega_0^4)  \bigg] \times \notag \\
&\exp[-\frac{1}{ \eta \omega_0}\lbrace H_0 -\alpha H'\rbrace]\\ 
&\textbf{(iv)} -\frac{1}{2}\frac{\hbar^2}{4}\partial^4_{p_y,y}=-\frac{\hbar^2}{2048 m^2 \pi^2 \eta^6 \omega_0^4}\bigg[ (-2p_y+m \omega_c x)^2-4 m \eta \omega_0(\omega_c^2(2p_x+m \omega_c y)^2-4m \omega_c^2 \eta \omega_0+ \notag \\
&8 m \omega_c y(2p_x+m \omega_c y)\omega_0^2-16m \eta \omega_0^3+16 m^2 y^2 \omega_0^4) \bigg] \exp[-\frac{1}{ \eta \omega_0}\lbrace H_0 -\alpha H'\rbrace] \\
&\textbf{(v)} -\frac{\hbar^2}{4}\partial^4_{p_y,y,p_x,x}W(x,y,p_x,p_y)=-\frac{\hbar^2}{4}\partial^4_{p_x,x,p_y,y}W(x,y,p_x,p_y)= \frac{\hbar^2}{2048 m^4 \pi^2 \eta^5 \omega_0^4}\bigg[\omega_c (-2 p_y +m \omega_c x)^2+ \notag \\
&4 m (-6 x^2 p_y \alpha+ 3 m \omega_c x^3 \alpha -\omega_c \eta)\omega_0)+4 m x (-2p_y+m \omega_c x)\omega_0^2\bigg]\bigg[\omega_c (2 p_x+m \omega_c y)^2-4 \eta m \omega_c \omega_0 + \notag \\
&4 m y (2 p_x +m \omega_c y)\omega_0^2\bigg] \exp[-\frac{1}{ \eta \omega_0}\lbrace H_0 -\alpha H'\rbrace] \label{term1}
\end{align}
We add these terms in the next step and arrange the anharmonicity terms separately to calculate the density operator using Eq.(\ref{wignerdensityrelation}):
\begin{align}
    &\rho_s=4\pi^2 \hbar^2 \bigg[1+\text{(H.M.O. terms)}+\alpha \bigg(\frac{3 i\hbar x^2(2p_x+m \omega_c y)}{64 m \pi^2 \eta^4 \omega_0^2}- \notag \\
   & \frac{3\hbar^2 x(2 p_x+m \omega_c y)^2(\omega_c x(-2 p_y+m \omega_c x)-4 \eta \omega_0+4 m \omega_0^2 x^2)}{256 m \pi^2 \eta^6 \omega_0^2}-....\bigg)+ \notag \\
   &\alpha^2 \bigg(\frac{9 \hbar^2 x^4((2p_x+m\omega_c y)^2-4 m \eta \omega_0)}{128\pi^2 \eta^6}\bigg)+....\bigg] \exp[-\frac{1}{ \eta \omega_0}\lbrace H_0 -\alpha H'\rbrace] \label{rhos}
\end{align}
Now we calculate $\ln(\rho_s)$ from Eq.(\ref{rhos}) using the  Baker-Campbell-Hausdroff formula and retain only upto first order in the expansion. Then we derive the expectation value of $\ln(\rho_s)$ in a mixed harmonic oscillator state (as anharmonicity is taken as a perturbation): 
\begin{align}
   S_{VN}=-k_B\langle \ln\rho_s \rangle=-k_B Tr(\rho \ln\rho) =-k_B \sum_j\langle j|\rho \ln \rho | j \rangle
\end{align}
The explicit derivation shows that the first order in $\alpha$ terms vanish and so the lowest order term $\alpha^2$ is retained as the anharmonicity parameter affecting $\ln(\rho_s)$ and the von-Neumann entropy of the system. Furthermore, in the high temperature classical limit, the diffusion is increased, resulting in the enhancement of the determinant $\eta$. Hence, the terms with higher orders of $1/\eta$ become negligibly small. So we keep the highest order term in $1/\eta$ to get the von-Neumann entropy as:
\begin{align}
  S_{VN}=(S_{VN})_{HM} -\frac{9 k_B\hbar^4 \alpha^2}{32 \eta^{6}} \sum_j \langle j|\frac{9k_B m\hbar^4 \omega_0\alpha^2}{8 \eta^{5}}   x^4   |j \rangle \label{lnrhofinal}
\end{align}
We calculate the \textit{Trace} in Eq.(\ref{lnrhofinal}), to get the final form of the $S_{VN}$ as:
\begin{align}
    S_{VN}=(S_{VN})_{HM}+\frac{9 k_B\hbar^6 \alpha^2}{32 m \omega_0 \eta^5}\left[(2 n_x+1)^2+2(n_x^2+n_x+1) \right]
\end{align}
where, $(S_{VN})_{HM}$ denotes the von-Neumann entropy of the harmonic oscillator and $n_x$ represents the average of the number operator corresponding to harmonic oscillator states at the long time limit, as has been mentioned in the main text.
\end{widetext}
%
\end{document}